\documentclass[prb, preprint]{revtex4}
\usepackage{graphicx,subfigure}
\usepackage{epsfig}
\usepackage{rotating}

\newcommand{\ba}{{\bf a}}
\newcommand{\oa}{{\overline{\alpha}}}
\newcommand{\oai}{{\overline{\alpha_i}}}
\newcommand{\oaj}{{\overline{\alpha_j}}}
\newcommand{\oba}{{\overline{\ba}}}
\newcommand{\obai}{{\overline{\ba_i}}}
\newcommand{\obaj}{{\overline{\ba_j}}}
\newcommand{\hba}{{\hat{\ba}}}
\newcommand{\hbai}{{\hat{\ba}_i}}
\newcommand{\hbaj}{{\hat{\ba}_j}}
\newcommand{\ohba}{{\overline{\hat{\ba}}}}
\newcommand{\ohbai}{{\overline{\hat{\ba}_i}}}
\newcommand{\ohbaj}{{\overline{\hat{\ba}_j}}}

\begin{document}

\title{Backbone and Side-chain Ordering in a Small Protein}

\author{Yanjie Wei} 
\affiliation{ Department of Physics, Michigan Technological University, 
             Houghton, MI 49931, USA}

\author{Walter Nadler}
\affiliation{Department of Physics, Michigan Technological University, 
             Houghton, MI 49931, USA}
\affiliation{ John-von-Neumann Institute for Computing, 
             Forschungszentrum J\"ulich, D-52425 J\"ulich, Germany}

\author{Ulrich H.E. Hansmann} 
\affiliation{ Department of Physics, Michigan Technological University, 
             Houghton, MI 49931, USA  }
\affiliation{ John-von-Neumann Institute for Computing, 
             Forschungszentrum J\"ulich, D-52425 J\"ulich, Germany}

\date{\today}
\maketitle

\newpage

\begin{center}
{\bf \Large Abstract}
\vspace{0.5cm}
\end{center}

We investigate the relation between backbone and side-chain ordering in a small protein. For this purpose we have performed multicanonical simulations of the villin headpiece subdomain HP-36, an often used toy model in protein studies. 
Concepts of circular statistics are introduced to analyze side-chain fluctuations.
In contrast to earlier studies on homopolypeptides (Wei et al., J. Phys. Chem. B, 111 (2007) 4244) we do not find collective effects leading to a separate transition. 
Rather, side-chain ordering is spread over a wide temperature range.
Our results indicate a thermal hierarchy of ordering events, with side-chain
ordering appearing at temperatures below the helix-coil transition but above the
folding transition. We conjecture that this thermal hierarchy reflects an underlying temporal order, 
and that side-chain ordering facilitates the search for the correct backbone topology.


\newpage

\section{Introduction}

The process by which a protein folds into its biologically active state cannot
be traced in all details solely by experiments. Fortunately, modern simulation
techniques have opened another window, often leading to new insight into
the dynamics and thermodynamics of folding \cite{Scheraga, Kolinski,Irbaeck,Wei,Skolnick,Wenzel}.
 Generalized ensemble techniques \cite{HO98} such as
parallel tempering \cite{PT1,PT2,H97f} or  multicanonical sampling
\cite{MU,MU2}, first introduced to protein science in Ref.~\onlinecite{HO},
 have made it possible to study the folding of small proteins 
(with  up to $\approx 50$ residues\cite{KH05}) {\it in silico}. 
Of particular interest is whether there are different distinct
 transitions in the folding process, and what their thermal order and relation is.

An example is the role of side-chain ordering. In recent studies on homopolymers~\cite{WNH06,WNH07a}, we found for certain amino acids a de-coupling of backbone and side-chain ordering. 
The ordering did not depend on the details of the environment, i.e. whether the molecules were in gas phase or solvent, but solely  on the particular side groups. 
It exhibited a transition-like character, marked by an accompanying peak in the specific heat. 
In the present work we  extend this study to proteins, i.e. heteropolymers of amino acids.  
 
 Our test protein is  the villin head piece subdomain HP-36 with which 
 we are familiar from earlier work \cite{ELP, HP36, TTH06}. This molecule 
 has raised considerable interest in
computational biology \cite{Freed,Ripoli}  as it is one of the smallest
proteins (596 atoms) with well-defined secondary and tertiary structure~\cite{McKnight:96} but at the same time still accessible to simulations~\cite{Duan:98}. 
Its structure  was resolved by NMR analysis and  is shown in Fig.~\ref{1vii}
as it is available in the Protein Data Bank \cite{PDB} (PDB code 1vii).
We use multicanonical sampling  to study
the thermal behavior of the protein in aqueous solvent over a wide range of temperatures from one single simulation. Such an approach is well-suited to overcome the problem 
of ``slowness'' of side-chain ordering observed in canonical simulations~\cite{Shimada,Kussell}. 

We observe that side-chain ordering occurs  over a wide range in  temperatures 
below the helix-coil transition. Although we  do not find the collective effects leading
 to a separate side-chain ordering transition that were observed for 
 homopolymers ~\cite{WNH06,WNH07a}, this result  indicates that   secondary 
 structure formation  is a necessary precursor for side-chain ordering. On the
 other hand, side-chain ordering occurs at higher  temperatures than those at which the protein backbone assumes its native fold. We conjecture that HP-36
 folds in a multi-step process,
 with side-chain ordering facilitating the search for the correct backbone topology.

\section{Methods}
Our simulations utilize  the ECEPP/2 force field \cite{EC} as
implemented in the 2005 version of the program package SMMP \cite{SMMP,SMMP05}.
 Here the interactions between the atoms of the protein are approximated
 by a sum $E_{ECEPP/2}$ consisting of electrostatic energy $E_C$, a  Lennard-Jones term $E_{LJ}$,
 hydrogen-bonding term $E_{HB}$ and a torsion energy $E_{Tor}$:
 \begin{eqnarray}
  E_{\text{ECEPP/2}} &=& E_C + E_{LJ}  + E_{HB} + E_{Tor} \nonumber \\
  &=&  \sum_{(i,j)} \frac{332 q_i q_j}{\epsilon r_{ij}} \nonumber \\
 &&   + \sum_{(i,j)} \left( \frac{A_{ij}}{r_{ij}^{12}} - \frac{B_{ij}}{r_{ij}^6} \right) \nonumber \\
 &&   + \sum_{(i,j)} \left( \frac{C_{ij}}{r_{ij}^{12}} - \frac{D_{ij}}{r_{ij}^{10}} \right) \nonumber \\
  && + \sum_l U_l ( 1\pm \cos(n_l \xi_l)) \;,
\label{energy}
\end{eqnarray}
where $r_{ij}$ is the distance between the atoms $i$ and $j$,  $\xi_l$ is the $l$-th torsion
angle, and energies are measured in kcal/mol.
The protein-solvent interactions are approximated by a solvent accessible surface term
\begin{equation}
  E_{solv} = \sum_i \sigma_i A_i \;.
\label{solventEnergy}
\end{equation}
The sum is over the  solvent accessible areas $A_i$ of all atoms $i$ weighted by solvation
parameters $\sigma_i$ as determined in Ref.~\onlinecite{OONS}, a common choice when the
ECEPP/2 force field is utilized.
Our previous experiences \cite{PTH,TTH06} have shown that $E_{solv}$ reproduces the  effects of protein-water interaction  {\it qualitatively} correct.
However, the temperature scale is often distorted, leading, for instance, to transitions at temperatures where water would be vaporized in nature. This problem can be remedied, however, by renormalization of the temperature scale upon comparison with experiments.

The above defined energy function leads  to a landscape that is characterized by a multitude of minima separated by high barriers. As the probability to cross an energy barrier of height $\Delta E$ is given by $\exp (-\Delta E/k_BT)$, $k_B$ being the Boltzmann constant, it follows that  extremely long runs are necessary to obtain sufficient statistics in regular canonical simulations at low temperatures. Hence, in order to enhance sampling we rely on the multicanonical approach \cite{MU,MU2} as described in Ref.~\onlinecite{HO}.  
Here, configurations are weighted with a non-canonical term $w_{MU} (E)$ usually determined iteratively to optimize certain properties of the simulation.
Thermodynamic averages 
of an observable $<O>$ at temperature $T$ are obtained by re-weighting \cite{FS}:
\begin{equation}
 <O>(T) =
         \frac{\int dx \; O(x) e^{-E(x)/k_BT} / w_{MU}[E(x)]} {\int dx \; e^{-E(x)/k_BT} / w_{MU}[E(x)]}
         \label{eq:reweighting}
 \end{equation}
 where $x$ counts the configurations of the system. 
 
 Most often the multicanonical weight is determined such that the probability distribution obeys
\begin{equation} 
   P_{MU}(E) \propto n(E) w_{MU}(E) \approx const~,
\end{equation}
where $n(E)$ is the spectral density of the system. However, in our implementation we do not require a constant histogram
but that the number of round-trips
 $n_{rt}$ between two  pre-set low and high energy values $E_{low}$ and $E_{high}$ 
 is maximal. $E_{high}$ is a an energy value typical for an disordered high temperature 
 state (in our example $E_{high}$ = -133.5 kcal/mol) while $E_{low} = -357$ Kcal/mol
 was chosen to correspond to  typical low-energy states as determined by us in preliminary studies. 
Obviously, the number of round-trips $n_{rt}$ between the lowest and highest temperature, 
$E_{low}$ and $E_{high}$, respectively, is a lower bound for the  statistically independent visits at 
the low energy states, and therefore a good measure for the efficiency of the simulation.
For this reason, it is desirable to maximize the number of round trips by optimizing 
$w_{MU}(E)$. This can be achieved in a  systematic way by the feedback algorithm 
described in Refs.~\onlinecite{THT, NH}. The resulting weights are given as supplemental material. 

A simulation of five million Monte Carlo sweeps (each consisting of 
 217 Metropolis steps that try to update all 217 dihedral angles of the molecule once) leads to  35 tunneling events, i.e. at least 35 independent configurations with energies smaller 
 than $-357$ kcal/mol. Every ten sweeps, we measure the 
energy $E$ with its respective contributions from Eq.~(\ref{energy}) and 
from the protein-solvent interaction energy $E_{solv}$. 
 Other quantities measured  are the radius of gyration $R_{gy}$ as a 
measure of the geometrical size, and
the number of helical residues $n_H$, i.e. residues where the pair of dihedral angles $(\phi,\psi)$ takes  
values in the range
  ($ -70^\circ \pm 30^\circ $, $ -37^\circ \pm 30^\circ $) \cite{Okamoto1995}.
Also we monitor the $RMSD$ (root mean square deviation) of various subsets of heavy atoms (backbone, sidechain, all) 
from the PDB structure.

Finally, all the 217 dihedral angles are recorded for later analysis of 
 their fluctuations and correlations. As the statistical analysis of dihedral angles has subtle pitfalls,  we present and justify our approach in the Appendix.

\section{Results and Discussions}

Multicanonical simulations allow determining
 thermodynamic quantities over a wide range of temperatures. The thermal evolution of the specific heat, for example, 
\begin{equation}
 C(T) = \frac{d}{dT} E = k_B \beta^2 \left (<E^2> - <E>^2\right)
 \end{equation}
 provides information about the temperatures where the protein changes
 its state. In earlier investigations \cite{WNH06, WNH07a} of homopolymers we observed  two separate peaks in the specific heat for particular amino acids, characterizing two well-defined transitions.  One peak was associated with a helix-coil transition, i.e. the ordering of the protein backbone. 
 The second peak, at a much lower temperature, could be related to an ordering of side chains.
These results indicated a two-step folding process upon lowering the temperature, starting with backbone ordering followed by side-chain ordering. 
How does the situation look like for a heteropolymer such as HP-36? 
 
 The specific heat curve in  Fig.~\ref{spec.heat} has only one marked peak at  $T= 505 \pm 8$ K;
  but it  also exhibits a shoulder around T=$300$K.
 As for the homopolymers, the peak in the specific heat  can be related to a helix-coil transition. 
 This interpretation is supported by  the inset where we
  display the average number $<n_H>$ of residues that are part of an $\alpha$-helix as a 
  function of temperature. The steep
 increase in this quantity at $T= 505$K is clearly correlated with the peak in the specific heat. 

 However, for HP-36 
 backbone ordering is more than just the formation of secondary structure. 
 The average radius of gyration $<R_{gy}>$, a measure for the compactness of a protein configuration, 
 as a function of temperature, is displayed in Fig.~\ref{rgy}. It indicates that backbone ordering 
  occurs in more than one step.
 Below $T=505$ K, most protein configurations have a high helix propensity. 
 Lowering the temperature further,
 compact structures become finally more frequent than 
 extended configurations with equal or even higher helicity. 
This two-step process in the backbone ordering can also be seen in the
 inset which displays the fraction of configurations with a $RMSD$ smaller than 6 \AA, 
 i.e. those that should fall within the free energy basin of the native structure~\cite{ScheragaOrSo}. 
  Final compactification and
   transition to nativeness are therefore concomitant processes.
We believe that the shoulder in the specific heat is correlated with  that final backbone ordering since it occurs close to the steepest parts of the decrease of $<R_{gy}>$ and 
the increase of nativeness.

 Hence, our results so far indicate a two-step process, but
 one that involves only the backbone. The first step, correlated with a critical temperature
 $T= 505$ K involves the formation of helical segments. In a second step, these arrange themselves
 to compact and native-like structures. The energy gain here is much smaller, and therefore this second  ordering step is observed at lower temperatures only. 

How does side-chain ordering fit into this picture? The behavior of
 the specific heat does not give any  indications of a separate transition related to 
 side-chain ordering. Such a transition could still exist --- albeit not associated 
 with large energy fluctuations. 
A quantity that describes side-chain ordering in a very general way is  
the average of the fluctuations of dihedral angles. We have calculated this 
quantity as described  in the appendix for  buried side chains
and compare it with fluctuations of angles belonging to side chains at the surface of the molecule. 
Both quantities are displayed in Fig.~\ref{buried} for all angles of a side chain, 
and in the inset solely for  the $\chi_1$ angle. 
For the buried residues one observes a single step ordering of the side chains. 
Immediately below the helix-coil transition the fluctuations decrease, 
indicating that here the formation of helical segments leads  already to some ordering of side chains. 
In the temperature range $300-500$~K the fluctuations decrease further, albeit less dramatic. This range corresponds to the shoulder in the specific heat 
and marks compactification and the formation of the tertiary backbone structure. 
 Residues at the surface exhibit a much smaller decrease of
fluctuations associated with the formation of helical segments.

At higher temperatures, side-chain ordering is  restricted to residues in the interior of a protein. 
 This is reasonable as here the side chain positions are more constraint by the geometry of the molecule. 
 For this reason, we have focused our further analysis on side chain angles of residues in the interior of the molecule.  
 Fig.~\ref{phenyl} shows the fluctuations of the $\chi_1$ angle for the
 residues Phe$_7$, Phe$_{11}$ and Phe$_{18}$. Fluctuations
 of these angles decrease 
 strongly over a small range of temperatures below the formation of the helical segments. 
 We note that Phe$_7$  exhibits ordering at a somewhat higher temperature than Phe$_{11}$ and Phe$_{18}$.
  
 The decrease in fluctuations is only loosely related to an increase in correlations between the $\chi_1$ angles of these three residues, see Fig.~\ref{corr}, where the data were determined as described in the appendix.
Phe$_7$ exhibits correlated fluctuations with Phe$_{11}$ already 
close to the helix coil transition. They persist and increase finally in the low temperature phase. Phe$_7$ and Phe$_{18}$ exhibit (anti-)correlations only below $350$~K. The most dramatic change occurs with Phe$_{11}$ and Phe$_{18}$: Their correlations start to occur around $450$ K, i.e. just when those angles are ordering; however, upon lowering the temperature the correlations switch to anti-correlations and increase in magnitude.

Note, that all correlated fluctuations of these side chains exhibit their steepest change  below $350$~K, where  Fig.~\ref{rgy} and its inset indicate the folding transition into the native backbone topology. On the other hand, this is the regime where angle fluctuations have subsided already.
Hence, for HP-36  the correct ordering of the side chains seem to predate tertiary structure formation. These results also indicate that the final arrangement of the side chains occurs collectively. 

The above  results indicate the following sequence of events in the folding of villin headpiece subdomain HP-36 upon lowering the temperature.
The first stage is the formation of helical segments, connected with a large gain in potential energy. 
Below this helix-coil transition is a large intermediate temperature range where  various 
 helical configurations  other than the native one 
 dominate for entropic reasons. This temperature 
range is also characterized  by an increased side-chain ordering that is more pronounced
 for side chains of residues in the interior that arrange themselves in coordinated way.
The heterogeneity of the sequence seems to destroy 
the phase transition-like character of side-chain ordering that was observed by us for
some homopolymers. Instead, the ordering is more gradual.  Only 
at temperatures below side-chain ordering, 
and connected with a much smaller gain
in energy than at the helix-coil transition, do the helical segments arrange themselves in native-like structures. 

Our results show a particular thermal order of the folding processes. It is natural to assume that this thermal order reflects a related temporal order of folding events. Hence, we conjecture that HP-36 folds in multi-step process where side chain and backbone ordering are interconnected. 
The initial step is the formation of helical segments. 
In a second step the protein collapses into more compact structures before it assumes its native state. 
This sequence of events is consistent with various computational~\cite{Lei,Pande,Mori} and experimental~\cite{Kubelka2003} studies  that also identify the formation of helical segments as the time limiting factor in the folding of HP-36. 
New is our observation that the search for the correct structure seems to be facilitated 
by the ordering of side chains subsequent to secondary structure formation. 
This scenario is also consistent  with recent mutagenesis experiments (relying on  nanosecond laser T-jump measurements) that emphasize the importance of  buried side chains for the rather short folding times of the villin headpiece~\cite{Kubelka2003,Kubelka2006}.  

\section{Summary and Outlook}

Choosing a well-studied small protein, the villin headpiece subdomain HP-36, 
we have presented methods that allow us to simulated and analyze ordering processes  taking place on the level of the side chain dihedral angles as well as at the level of the backbone structures.
Our results indicate a thermal hierarchy of ordering events with side-chain ordering appearing at temperatures below the helix-coil transition but above the folding transition. 
We believe that the observed thermal hierarchy of folding reflects an underlying temporal sequence of these ordering processes in actual protein folding dynamics. 
We conjecture that side-chain ordering facilitates the search for the correct backbone topology.
Further studies along these lines on different proteins will elucidate how general such a scenario is.

 \vspace{1cm}
{\em Acknowledgments }
Support by a research grant (CHE-0313618) of the National Science Foundation (USA) is acknowledged.

\begin{appendix}

\section{Statistical Analysis of Dihedral Angles}

Correct statistical analysis of dihedral angles is somewhat subtle because of their periodicity modulo $2\pi$. 
This property excludes the use of regular statistical measures like the mean angle, $\left<\alpha\right>$, 
or its variance, $\left<(\alpha -\left<\alpha\right>)^2\right>$. 
The reason is that the numerical values of those quantities depend on the reference frame chosen, 
 e.g. $\left[-\pi,\pi\right]$ or  $\left[0,2\pi\right]$ or any other interval of length $2\pi$.
Moreover, choosing an inappropriate reference frame can lead, e.g. to the spurious appearance of a bimodal distributions from an underlying unimodal one.

On the other hand, there exist the well-established mathematical fields of 
{\it circular} or 
{\it directional statistics}~\cite{Fisher1995,Mardia1999,Topics2001}
that deal with such problems. 
However, we believe that some the quantities and equations used there introduce unnecessary complications and do not fully reflect the underlying physical concepts. So, here we will borrow some ideas from that field, but we will not fully follow that approach. 

The important fundamental idea introduced in circular and directional statistics is that an angle $\alpha$ can be viewed as a two-dimensional vector of unit length
\begin{equation}
{\bf a} = { \cos(\alpha) \choose \sin(\alpha) } \quad ,
\end{equation} 
a concept that bears some similarity to, e.g., a spin in an XY-model~\cite{XYmodel} treated in statistical physics.
Consequently, we are interested in the mean direction $\oba$, also considered to be a unit vector.
It can be determined from the averaged vector
\begin{equation}
\left< {\bf a} \right> = { \left< \cos(\alpha) \right>
\choose \left< \sin(\alpha) \right> } 
\quad .
\label{eq:avba}
\end{equation} 
which is usually smaller than a unit vector,
\begin{equation}
R^2(\alpha)=\left<\cos(\alpha)\right>^2+\left<\sin(\alpha)\right> ^2 < 1
\quad , 
\label{eq:Rsquare}
\end{equation}
by
\begin{equation}
\oba=\frac{1}{R(\alpha)}  \left< {\bf a} \right> 
\label{eq:babar}
\end{equation}
From this mean direction vector a corresponding mean angle $\oa$ could be determined in an appropriate frame, 
\begin{equation}
\oba \equiv { \cos(\oa)\choose \sin(\oa)} \quad .
\end{equation}
Notice that - as we will see below - most often it is not necessary to determine that angle. Rather, it is sufficient to work with either the mean vector $\left<\ba\right>$, Eq.~(\ref{eq:avba}), or the mean direction vector $\oba$ , Eq.~(\ref{eq:babar}).

In this contribution we concentrate mostly on $fluctuations$ and $correlations$ between dihedral angles. Correlation analysis, in particular, is a somewhat complex field in the directional statistics literature, sometimes motivated and dominated by the fact that the underlying data are temporal and the goal is the detection of circadian rhythms~\cite{Fisher1995,Mardia1999}. Moreover, the quantities employed for describing fluctuations do not always match up with 
those employed for describing correlations.
Below we sketch the problems and justify our approach.

The simplest measure for fluctuations is based on the length of the average vector, Eq.~(\ref{eq:Rsquare}). The {\it circular variance} is given simply by 
\begin{equation}
V\left(\alpha\right) = 1- R\left(\alpha\right) \quad .
\label{eq:V0}
\end{equation}
$V=0$ corresponds to vanishing fluctuations, 
while $V=1$ describes the case of an equidistribution of angles over the full range, i.e. maximal fluctuations.
Interestingly, the circular variance can be derived, too,   
by considering the deviation vectors from the mean direction, i.e.
\begin{eqnarray}
V\left(\alpha\right) &=& 
\frac{1}{2} \left<\left|\ba-\oba\right|^2 \right> 
\nonumber \\
&=& \frac{1}{2} \left(
\left<\ba^2\right> - 2\left<\ba\right>\cdot\oba + \oba^2 \right) 
\nonumber \\
&=& 1-R(\alpha) 
\quad .
\label{eq:fluct}
\end{eqnarray}

Ideally, in order to systematically analyze correlations and fluctuations together, a covariance function $C\left(\alpha_i,\alpha_j\right)$ is necessary that generalizes the fluctuation measure employed. 
Combining the chosen covariance and fluctuation functions, the correlation matrix is finally given by
\begin{equation}
\rho\left(\alpha_i,\alpha_j\right) = 
\frac{C\left(\alpha_i,\alpha_j\right)}
{\sqrt{V(\alpha_i)V(\alpha_j)}} \quad .
\label{eq:corr}
\end{equation}
$\rho=0$ denotes vanishing correlations, either since there are no fluctuations at all, or because the fluctuations are uncorrelated. $\rho\to\pm 1$ corresponds to full correlation or anti-correlation of the fluctuations, respectively.

Unfortunately, a straightforward extension
from Eqs.~(\ref{eq:V0}) and (\ref{eq:fluct}), e.g. defining the covariance function as the scalar product of the respective deviation vectors from the mean direction,
$C\left(\alpha_i,\alpha_j\right) \propto
\left<\left(\ba_i-\obai\right) \cdot
      \left(\ba_j-\obaj\right)\right>$,
is not possible. This quantity  does not vanish if the angles are statistically independent, as it should for a proper covariance.
Instead, replacing the deviations from the mean direction by the deviations from the {\it mean vector} does result in a seemingly proper covariance function,
\begin{equation}
 C_{\rm diff}\left(\alpha_i,\alpha_j\right) =
\left<\left(\ba_i-\left<\ba_i\right>\right) \cdot
      \left(\ba_j-\left<\ba_j\right>\right)\right>
      \quad .
\label{eq:cov1}
\end{equation}
The related variance function differs from Eq.~(\ref{eq:V0}) though,
 \begin{eqnarray}
V_{\rm diff}\left(\alpha\right) &=& 
\left<\left|\ba-\left<\ba\right>\right|^2 \right> =
\left<\ba^2\right> -
\left<\ba\right>^2\nonumber \\
&=&
1-\left[
\left<\cos(\alpha)\right>^2+\left<\sin(\alpha)\right> ^2\right] 
\nonumber \\
&=& 1-R^2(\alpha) 
\quad .
\label{eq:V1}
 \end{eqnarray}
Both forms, $V\left(\alpha\right)$ and $V_{\rm diff}\left(\alpha\right)$, are related by a monotonic --- albeit nonlinear --- mapping and describe fluctuations in a qualitatively similar way. The only quantitative difference is that Eq.~(\ref{eq:V1}) better resolves the small fluctuation regime while Eq.~(\ref{eq:V0}) does that with the regime of large fluctuations. 

While we do not consider the changed variance to be a problem, there is one with Eq.~(\ref{eq:cov1}). Although $C_{\rm diff}\left(\alpha_i,\alpha_j\right)$ exhibits the correct behavior in the limit of statistical independence of the angles, we have observed that problems arise in the regime of larger correlations. This is due to the fact that $\left|\left<\ba_i\right>\right|\ne\left|\left<\ba_j\right>\right|$ usually holds, which leads to an imbalance in the treatment of the respective deviation vectors. 

The authors of Ref.~\onlinecite{Topics2001} suggest to describe correlations between angles by the covariance function
 \begin{equation}
C_{\sin}\left(\alpha_i,\alpha_j\right) = \left<\sin(\alpha_i-\oai)\sin(\alpha_j-\oaj)
\right> \quad ,
\label{eq:cov2}
\end{equation}
This function also exhibits the correct behavior for independently distributed angles, and - again - the related variance function differs from Eq.~(\ref{eq:V0}),
 \begin{equation}
V_{\sin}\left(\alpha\right) = \left<\sin^2(\alpha-\oa)\right> \quad .
\label{eq:V2}
\end{equation}
Notice that this measure of fluctuations necessarily includes
higher order moments of the angular trigonometric functions than those Eqs.~(\ref{eq:V0}) and (\ref{eq:V1}) use. Consequently, there does not exist a simple analytic mapping to the circular variance, and - particularly for large fluctuations - a non-monotonic relationship is possible~\cite{footnoteNonMonotonic}.

We note that, as mentioned above, it is actually not necessary to determine the average angle $\oa$ explicitly for evaluating Eq.~(\ref{eq:cov2}). 
Rather, this equation also has a vector representation, 
albeit by using the cross product of vectors in addition to the scalar product.
Extending $\ba$ to a 3d vector via
\begin{equation}
\hba=\left(
\begin{array}{c} 
 \cos(\alpha) \\ \sin(\alpha) \\ 0 \\
\end{array} 
 \right)
\quad ,
\label{eq:3dvector}
\end{equation}
and using trigonometric identities, it can be easily seen that the the sine of the angle difference is given by the $z$-component of the cross product
$(-\hba\times\ohba)$. Consequently, the covariance (\ref{eq:cov2}) can be represented as
\begin{equation}
C_{\sin}\left(\alpha_i,\alpha_j\right) =
\left<\left(\hbai\times\ohbai\right) \cdot
      \left(\hbaj\times\ohbaj\right) \right> \quad .
\label{eq:covVec}
\end{equation}
Analogously, the corresponding fluctuations are represented via
\begin{equation}
V_{\sin}(\alpha) =
\left<\left|\hbai\times\ohbai\right|^2 \right> \quad .
\label{eq:fluctHigher}
\end{equation}

As outlined above, we would have preferred to systematically analyze fluctuations and correlations together,
either using Eqs.~(\ref{eq:V1}) and (\ref{eq:cov1}), or Eqs.~(\ref{eq:V2}) and (\ref{eq:cov2}).
However, the problems with the covariance Eq.~(\ref{eq:cov1}) --- imbalance in the large correlations regime --- and the variance Eq.~(\ref{eq:V2}) --- non-monotonicity in the large fluctuations regime --- do not allow this. 

Rather, we decided to employ a hybrid approach:
When dealing with fluctuations we always rely on the circular variance, Eqs.~(\ref{eq:V0}) since it is the simplest reliable approach. 
When dealing with correlations, we use the covariance $C_{\sin}\left(\alpha_i,\alpha_j\right)$, Eq.(\ref{eq:cov2}). 
Necessarily, we have to employ the problematic variance $V_{sin}(\alpha)$, Eq.(\ref{eq:V2}), as normalization in determining the correlation function $\rho\left(\alpha_i,\alpha_j\right)$, Eq.~(\ref{eq:corr}).
Since in our case correlations arise only in the regime where fluctuations are small, we feel that is an acceptable approach. It also outweighs the problems that arise from using  Eq.~(\ref{eq:cov1}).

We emphasize in closing that --- to our knowledge --- no satisfying approach exists yet to treat strong  dihedral angle correlations in the large fluctuations regime.

\end{appendix}

\newpage

%
%
%
\clearpage
{\huge Figure captions:}

\begin{description}
\item{Fig.~1:} Structure of HP-36 (picture was obtained by VMD ~\cite{vmd} ).
\item{Fig.~2:} Specific heat as a function of temperature. 
The inset displays the helicity as a function of temperature.
\item{Fig.~3:} Radius of gyration as a function of temperature.
The inset shows  the fraction of configurations with a backbone RMSD from the PDB structure less than 6 \AA. 
\item{Fig.~4:} Averaged fluctuations, Eq.~(\ref{eq:V0}), of side chain angles from buried and surface side groups, respectively;  
 the inset shows average of only the $\chi_1$ angle fluctuations. Note the errorbars 
 denote the average of the errors in the fluctuations of each individual $\chi$ angle.
\item{Fig.~5:} Averaged fluctuations, Eq.~(\ref{eq:V0}),
 of  $\chi_1$  for Phe7, Phe11, and Phe18.
\item{Fig.~6:}
   Correlations, Eq.~(\ref{eq:corr}), based on Eqs.~(\ref{eq:cov2}) and (\ref{eq:V2}), of  $\chi_1$  fluctuations between Phe7 and Phe11, Phe7 and Phe18, and Phe11 and Phe18, respectively; see also the discussion in the appendix.
\end{description}

%


%
\setcounter{figure}{0}
%
\clearpage
\begin{sidewaysfigure}
    \includegraphics[width=1.0\columnwidth]{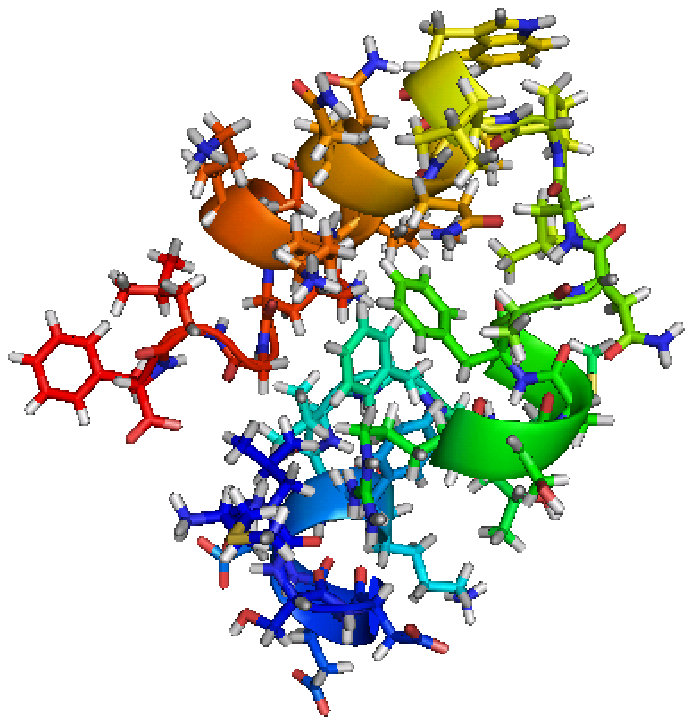}
\caption{\label{1vii}}
\end{sidewaysfigure}

\clearpage
\begin{sidewaysfigure}
    \includegraphics[width=1.0\columnwidth]{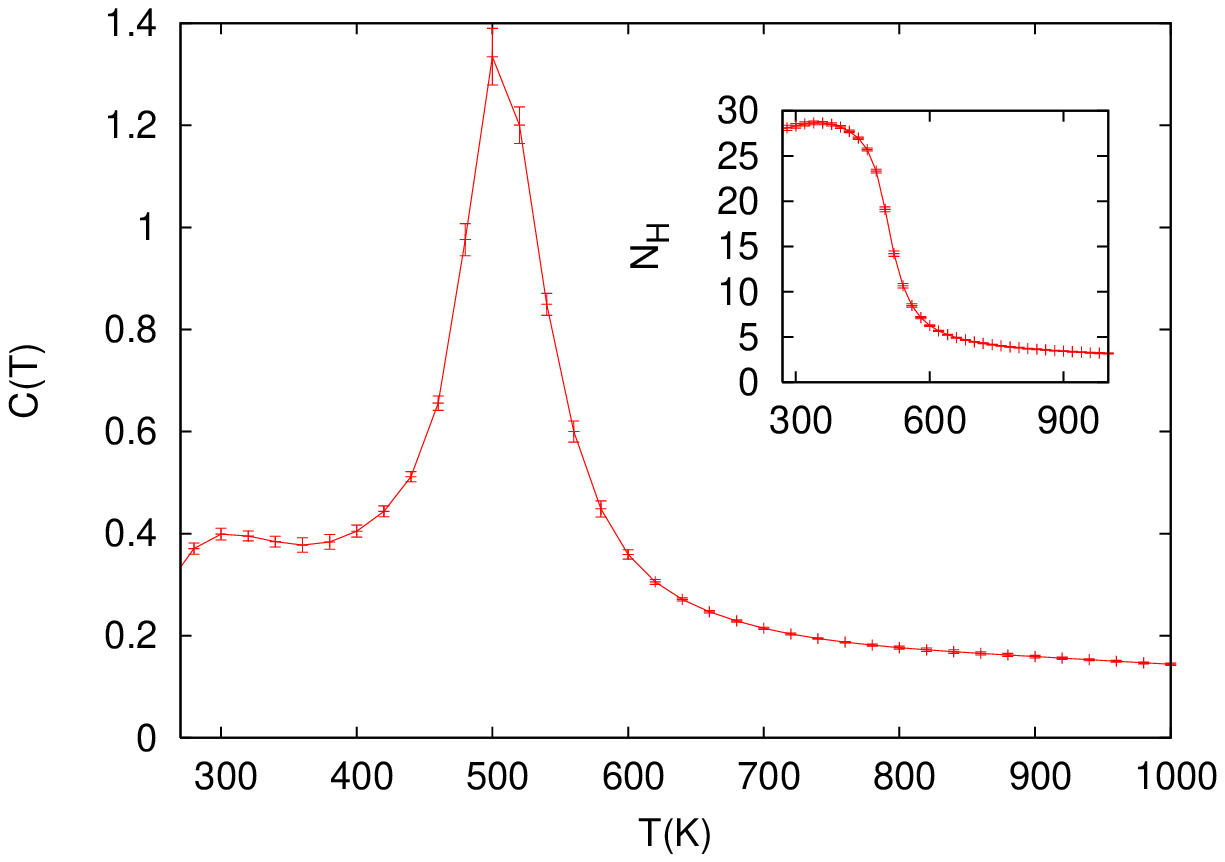}
\caption{\label{spec.heat}}
\end{sidewaysfigure}

\clearpage
\begin{sidewaysfigure}
    \includegraphics[width=1.0\columnwidth]{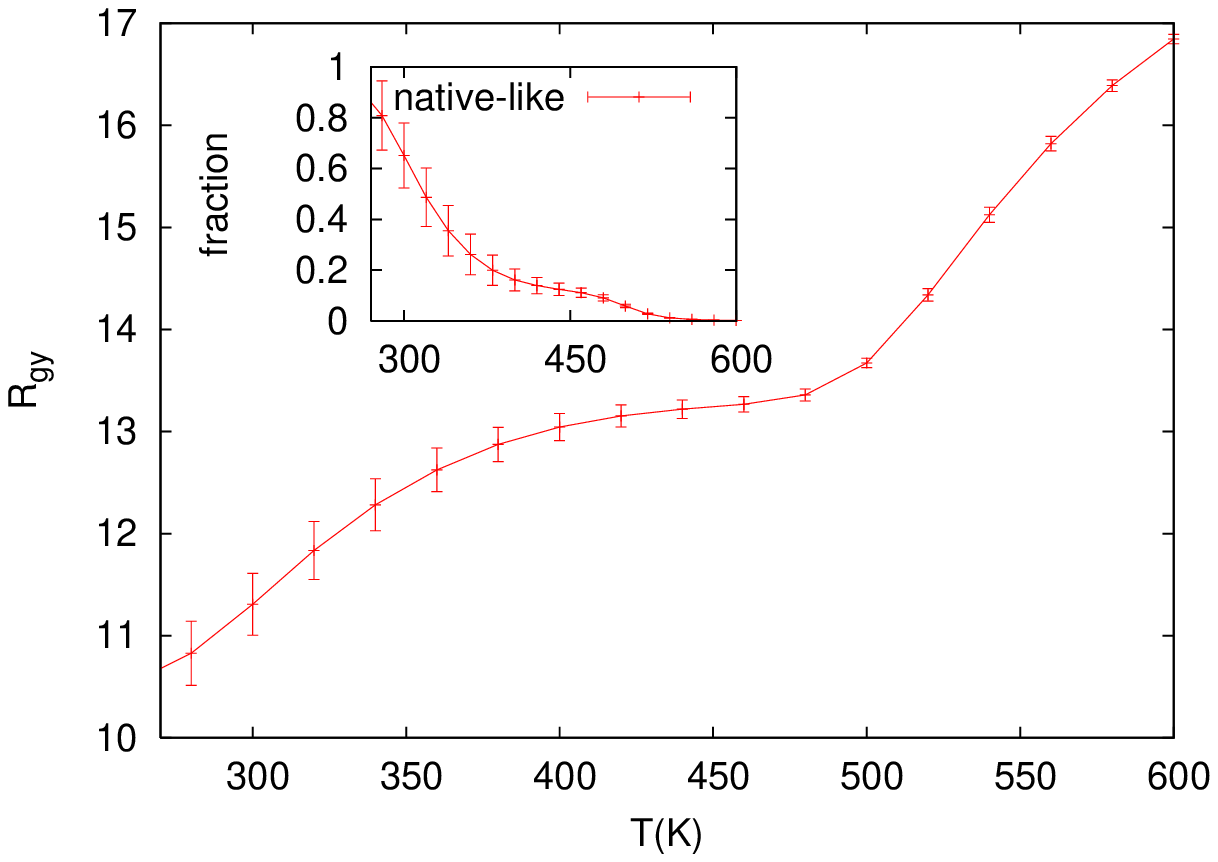}
\caption{\label{rgy}}
\end{sidewaysfigure}

\clearpage
\begin{sidewaysfigure}
    \includegraphics[width=1.0\columnwidth]{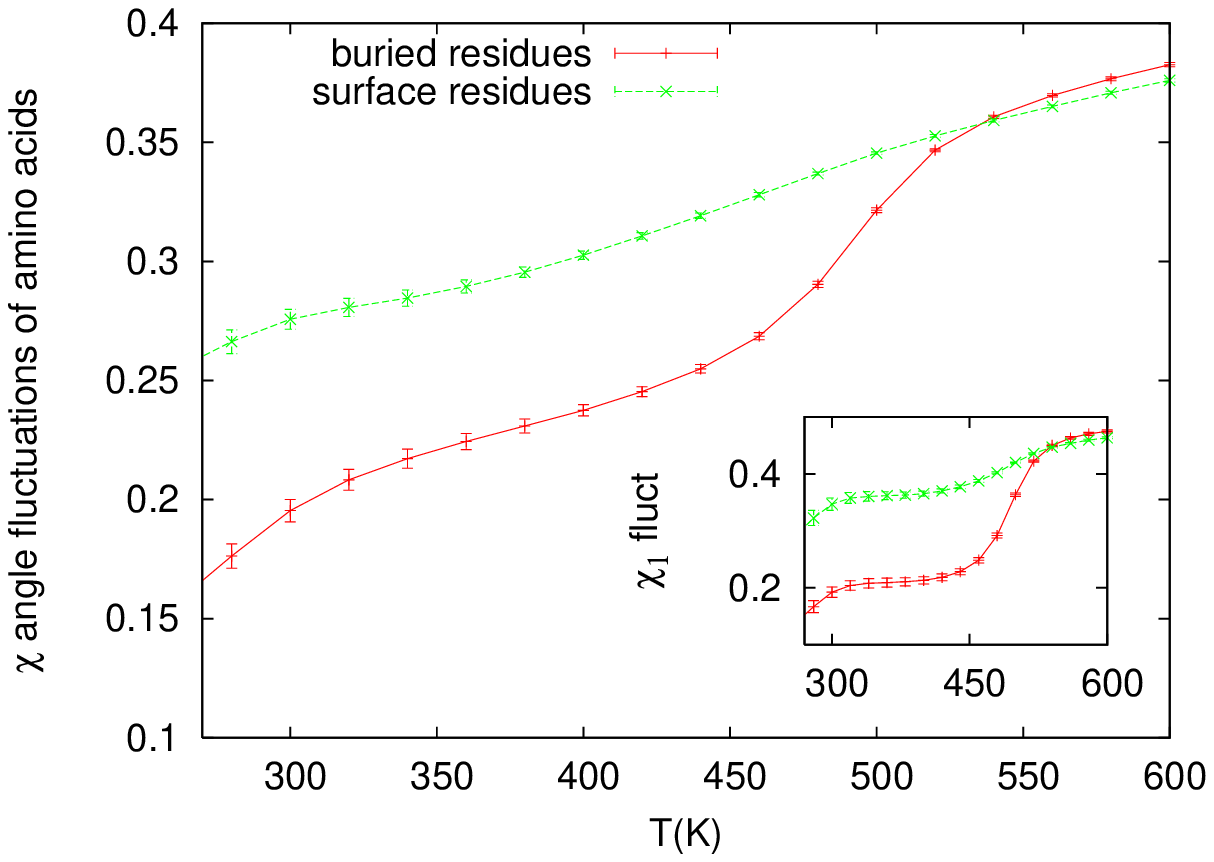}
\caption{\label{buried}}
\end{sidewaysfigure}

\clearpage
\begin{sidewaysfigure}
    \includegraphics[width=1.0\columnwidth]{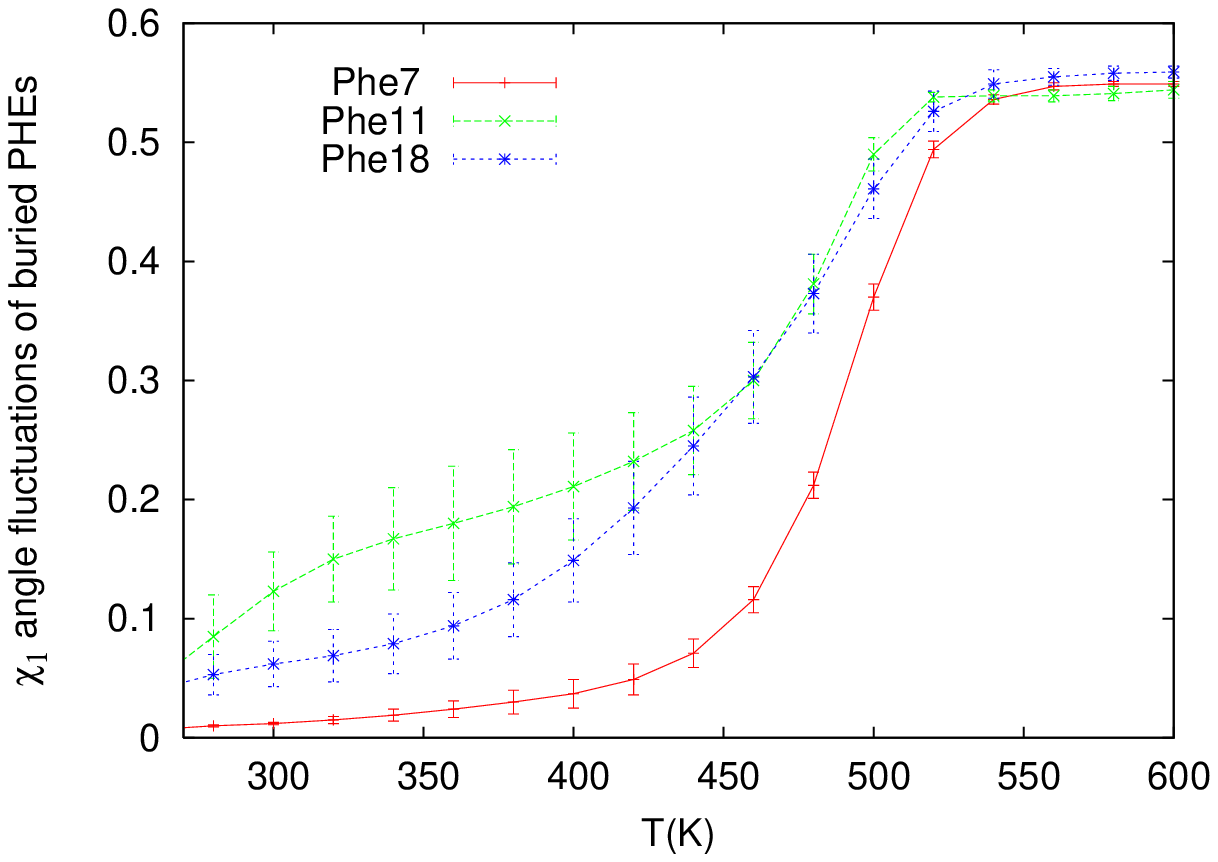}
\caption{\label{phenyl}}
\end{sidewaysfigure}

\clearpage
\begin{sidewaysfigure}
    \includegraphics[width=1.0\columnwidth]{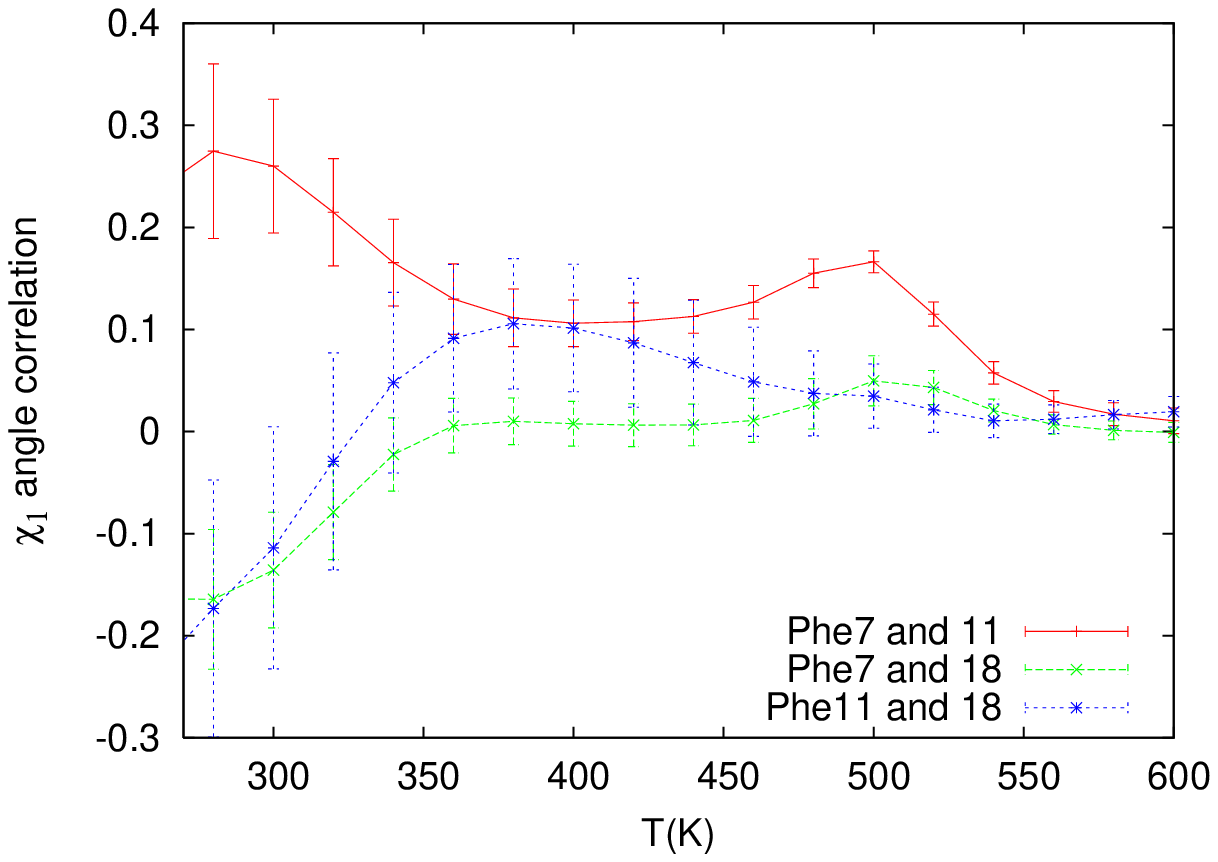}
\caption{\label{corr}}
\end{sidewaysfigure}


\begin{thebibliography}{99}
\bibitem{footnoteNonMonotonic} 
For example, for angles $\alpha$ distributed equally in the interval $\left[-\beta,\beta\right]$, the resulting variance is $V_{\sin}\left(\alpha\right)=\left[2\beta-\sin(2\beta)\right]/4\beta$.
 Clearly, larger fluctuations give rise to oscillations in $V_{\sin}\left(\alpha\right)$. Consequently, only the range $0\le V_{\sin}\left(\alpha\right) < 1/2$ describes fluctuations somewhat reliably, the latter value holding for an equidistribution of angles.
\end{thebibliography}
\end{document}